\documentclass{JHEP3}\newcommand{\format} {\JHEPformat}

\usepackage{amsmath}
\usepackage{mathrsfs}
\usepackage{epsfig}
\usepackage{latexsym}

\newcommand{\JHEPformat} {
\bibliographystyle{JHEP}
\newcommand{\maketitlepage} {}
\abstract{\theabstract}
\keywords{\thekeywords}
\preprint{\thepreprint}
}

\newcommand{\TITLE}[1] {\newcommand{\thetitle} {#1}\title{#1}}
\newcommand{\ABSTRACT}[1] {\newcommand{\theabstract} {#1}}

\newcommand{\ADDRESS}[1] {\newcommand{\theaddress} {#1}}
\newcommand{\DATE}[1] {\newcommand{\thedate} {#1}\date{#1}}
\newcommand{\KEYWORDS}[1] {\newcommand{\thekeywords} {#1}}
\newcommand{\PREPRINT}[1] {\newcommand{\thepreprint} {#1}}

\def\half{\frac{1}{2}}

\newcommand{\be}{\begin{equation}}
\newcommand{\ee}{\end{equation}}
\newcommand{\bea}{\begin{eqnarray}}
\newcommand{\eea}{\end{eqnarray}}
\TITLE{On the pp-wave limit and the BMN structure of new Sasaki-Einstein spaces}
\ADDRESS{School of Physics and Astronomy\\
  The Raymond and Beverly Sackler Faculty of Exact Sciences\\
  Tel Aviv University, Ramat Aviv, 69978, Israel
}

\author{Stanislav Kuperstein\\
Physique Th\'{e}orique et Math\'{e}matique, 
\\ Universit\'{e} Libre de Bruxelles and International Solvay Institutes, \\
ULB-Campus Plaine C.P.231, B-1050 Bruxelles, Belgium. \\
E-mail:
\email{skuperst@ulb.ac.be}
}

\author{Oded  Mintkevich, Jacob Sonnenschein\\
School of Physics and Astronomy,\\
The Raymond and Beverly Sackler Faculty of Exact Sciences,\\
Tel Aviv University, Ramat Aviv, 69978, Israel.\\
E-mail:
\email{oded@post.tau.ac.il, cobi@post.tau.ac.il}
}

\ABSTRACT{
We construct the \emph{pp}-wave string associated with the Penrose limit of 
$Y^{p,q}$ and $L^{p,q,r}$ families of Sasaki-Einstein geometries. 
We identify in the dual quiver gauge theories the chiral and the non-chiral operators that 
correspond to the ground state and the first excited states. 
We present an explicit identification in a prototype model of $L^{1,7,3}$. }

\DATE{September 2006}


\PREPRINT{TAUP-2835-06\\ ULB-TH/06-20 \\{\tt hep-th/0609194}} 

\format

\begin{document}

\maketitlepage

% ---------------------------------------------------------------------------------------------------

\section{Introduction}

During the last two years a series of papers has been published on new infinite families of $5$-dimensional
Sasaki-Einstein geometries $Y^{p,q}$ and $L^{p,q,r}$
\cite{Gauntlett:2004yd,Martelli:2004wu,Bertolini:2004xf, Benvenuti:2004dy, Cvetic:2005ft, Benvenuti:2005cz,Benvenuti:2005ja,Franco:2005sm,
Cvetic:2005vk,Butti:2005sw,Martelli:2005wy}. 
The quiver gauge theories (QGT) dual to these
backgrounds have been constructed explicitly and analyzed in detail.
The results of these papers change the status quo in the gauge/gravity duality, since
until recently the only non-trivial superconformal QGT in the context 
of AdS/CFT was provided by Klebanov and Witten (KW) \cite{Klebanov:1998hh}. The supergravity dual of this model 
is $T^{1,1}$, which appears now to be a special case of the $Y^{p,q}$ family.

According to the original Maldacena conjecture
the chiral operators of the strongly coupled $\mathcal{N}=4$
$SU(N)$ gauge theory are in one-to-one correspondence with the modes of type IIB supergravity
on $AdS_5 \times S^5$  \cite{Aharony:1999ti}. The precise form of the map, however, remains a mystery.
One of the main breakthroughs in the study of the correspondence was the idea to consider
only states with very large angular momentum along the equator of $S^5$ \cite{Berenstein:2002jq}.
This amounts, effectively, to taking
the Penrose limit of $AdS_5 \times S^5$. This limit results in a maximally supersymmetric 
$pp$-wave background \cite{Gueven:2000ru,Blau:2001ne,Blau:2002mw,Blau:2002dy}. 
Remarkably the string theory in this background is exactly solvable
in  the light-cone gauge \cite{Metsaev:2001bj,Metsaev:2002re}.
Combined with the AdS/CFT duality this provides an explicit 
relation between the dimension and the $R$-charge of gauge theory operators dual to the 
string excitations.  These single trace operators with high $R$-charge are 
known as the BMN operators \cite{Berenstein:2002jq}.

It appears that  for an arbitrary Sasaki-Einstein space $M_5$, 
the Penrose limit around an appropriate null geodesic on $AdS_5 \times M_5$ results in a 
maximally supersymmetric $pp$-wave background \cite{Itzhaki:2002kh}. 
In particular, this can be done for the conifold
background. It implies that, like in the $\mathcal{N}=4$ case, apart from a BMN operator
corresponding to the 
string ground state, there are eight additional BMN operators dual to to the degenerate first excited state.
In the papers \cite{Itzhaki:2002kh,Gomis:2002km,PandoZayas:2002rx} 
these operators were constructed explicitly in terms of the chiral fields of the KW model. 
In short, four operators are built by acting with space-time derivatives on the BMN operator
dual to the ground state, two additional operators are constructed from the chiral fields 
in a way similar to the ground state operator, while the last two BMN operators are built from
the two $SU(2)$ currents of the gauge theory. These operators are non-chiral, but still have
protected quantum numbers, as can be verified from the supergravity spectroscopy analysis.

Similar analysis was also carried out in \cite{Gimon:2002nr} for the Klebanov-Strassler 
\cite{Klebanov:2000hb} and
Mal-dacena-N\'u\~nez \cite{Maldacena:2000yy} backgrounds, which are dual to 
non-conformal $\mathcal{N}=1$ supersymmetric gauge theories. Like in the conformal cases the Penrose
limits around null geodesics located in the IR region yield exactly solvable string theory models.
These represent the non-relativistic motion and low-lying excitations of heavy hadrons with mass proportional to 
a large global charge. It was further shown in \cite{Gimon:2002nr} that these hadrons, also termed ``annulons",
take the form of heavy non-relativistic strings\footnote{
See \cite{Kuperstein:2003yt} and \cite{Apreda:2003gs} for the analogous discussion 
of the non-supersymmetric deformation of the  Klebanov-Strassler background. 
The ``annulons" of the Maldacena-N\'u\~nez background and its non-supersymmetric version 
appear in \cite{Apreda:2003gs,Bigazzi:2004yt}
For other confining backgrounds see \cite{Bertoldi:2004rn, Bigazzi:2004ze}.
See also \cite{Cotrone:2006re} for a general discussion of ``annulons" in a confining gauge theory 
admitting a supergravity dual background.}.

In our paper we take a step further. We take the Penrose limit of the $Y^{p,q}$ and $L^{p,q,r}$ 
backgrounds and analyze the BMN operators of the dual gauge theories.
In our analysis we make extensive use of the underlying 
K\"ahler quotient structure of the CY cones. 
It proves to be a very powerful tool for the construction and classification 
of the chiral gauge invariant operators.  
We identify the ground state dual operator as well as six chiral BMN operators corresponding 
to the first excited state in the $Y^{p,q}$ and the $L^{p,q,r}$ cases.

Exactly like in the conifold case there are two non-chiral operators dual
to the first excited string states. Note, however, that the $Y^{p,q}$ geometries have only one
$SU(2)$ isometry factor, while the $L^{p,q,r}$ spaces have no $SU(2)$ isometry at all.
We therefore cannot built the two non-chiral BMN operators entirely from the $SU(2)$ currents like in the
$T^{1,1}$ case.
This problem was first addressed by \cite{Benvenuti:2005cz}, where the so called ``short-cut" non-chiral 
operator was constructed for the $Y^{p,q}$ case. Although this non-chiral operator is not a component of any 
current, it seems to have the right quantum numbers matching the first excited string state.
In the $L^{p,q,r}$  case there are two independent "short-cuts". In this paper we give
a general idea how to build these operators for a general $L^{p,q,r}$ theory and perform 
an explicit construction for the $L^{1,7,3}$ special case.

The outline of the paper is as follows.
In Section \ref{Ypq} we show that the Penrose limit of the $AdS_5 \times Y^{p,q}$ background
yields the maximally supersymmetric $pp$-wave metric. We also rewrite the light-cone Hamiltonian
in terms of the currents and the conformal dimension operator of the dual gauge theory.
In Section \ref{FT} the BMN construction \cite{Itzhaki:2002kh,Gomis:2002km,PandoZayas:2002rx}
for the KW model is briefly reviewed.
We then rewrite the light-cone Hamiltonian it terms of the derivatives with respect to the 
K\"ahler quotient variables of the conifold and reproduce the results of 
\cite{Itzhaki:2002kh,Gomis:2002km,PandoZayas:2002rx}. 
This  method is further used 
to reconstruct the chiral BMN operators of the $Y^{p,q}$ theory. 
We also comment on the short-cut operator of \cite{Benvenuti:2005cz}.
Section \ref{Lpqr} is devoted to the Penrose limit of the $L^{p,q,r}$ backgrounds.
Rather than working with the gauge theory fields, we work again with the K\"ahler quotient coordinates, 
successfully constructing the chiral BMN operators. We end this section with a comment on the ``short-cut"
operators. In Section \ref{L1735} we work out the $L^{1,7,3}$ example providing
an explicit construction of the chiral and non-chiral ``short-cut" operators.
We close in Section \ref{conc}  with some remarks and
suggestions for further research.

%-----------------------------------------------------------------------------
\section{The Penrose limit of Sasaki-Einstein $Y^{p,q}$ spaces}

\label{Ypq}

In this section we will construct a maximally supersymmetric \emph{pp}-wave background 
by taking a Penrose limit of the AdS${}_5 \times Y^{p,q}$ supergravity solution.
The global AdS${}_5$ metric is:

\be
\frac{1}{R^2} ds^2_{AdS_5} = - dt^2 \cosh^2 \rho + d \rho^2 + \sinh^2 \rho d \Omega_3^2.
\ee 
Let us now briefly review the geometry of the Sasaki-Einstein metric
on $Y^{p,q}$. It is given by \cite{Gauntlett:2004yd, Martelli:2004wu}:

\bea \label{eq:5d}
\frac{1}{R^2} ds^2_{Y^{p,q}} &=& \frac{1-c y}{6} \left( d \theta^2 + \sin^2  \theta d \phi^2 \right) +
    \frac{dy^2}{H(y)} + \frac{H(y)}{36} \left( d \beta + c \cos \theta d \phi \right)^2 + \\
\nonumber        
   && + \frac{1}{9} \left( d {\psi^\prime} - \cos \theta d \phi + y \left( d \beta + c \cos \theta d \phi \right) \right)^2. 
\eea
where 

\be
H(y)= 2 \frac{a-3y^2+2 c y^3}{1-c y}.
\ee
The conifold case corresponds to $c=0$. Otherwise one can re-scale the coordinate $y$
to put $c=1$. 
Written in this way 
the first line of (\ref{eq:5d}) corresponds
to the $4$d K\"ahler-Einstein basis parameterized by the coordinates $\theta, \phi, \beta$ and $y$,
while the second line is associated with the $U(1)$-fibration parameterized by the angle $\psi^\prime$. 
The coordinates $\theta, \phi$ and $y$ span the range:

\be
0 \le \theta \le \pi,  \qquad 0 \le \phi  < 2 \pi
\qquad \textrm{and} \qquad y_1 \le y \le y_2,
\ee
where the constants $y_{1,2}$ are determined by:

\be
y_{1,2} = \frac{1}{4 p} \left( 2 p \mp 3 q - \sqrt{4p^2-3q^2} \right).
\ee
To see the periods of $\beta$ and $\psi^\prime$
one has to use angles $\alpha$ and $\psi$ defined by:

\be \label{eq:alpha_psi}
\alpha = -\frac{1}{6} ( \beta + c \psi^\prime )
\qquad \textrm{and} \qquad \psi = \psi^\prime.
\ee
In these coordinates:

\be
0 \le \alpha < 2 \pi \ell,  \qquad 0 \le \psi < 2 \pi
\ee
with 

\be
\ell = \frac{q}{3 q^2 -2p^2+p\sqrt{4p^2-3q^2}}.
\ee
The conifold case corresponds to $p=1$, $q=0$ with $\ell=1/3$.
For $p>1$, $q=0$ the metric describes the orbifold of the conifold $T^{1,1}/\mathcal{Z}_p$
and $\ell=(3p)^{-1}$.
The $5d$ compact space $Y^{p,q}$ has $SU(2) \times U(1)_F \times U(1)_R$ isometry
and its local structure is identical for any $p,q$.
The only impact of the $p$ and $q$ parameters is on the periodicity of the angular coordinate $\alpha$.
The $SU(2)$ isometry becomes explicit when the coordinates (\ref{eq:alpha_psi})
are used. In this case one can conveniently rewrite the $4d$ K\"ahler-Einstein metric
in terms of the $SU(2)$ left-invariant Maurer-Cartan forms $\sigma_{i=1,2,3}$
built from the angles $\theta,\phi$ and $\psi$.
The Killing-Reeb vector $2 i \partial_{\psi^\prime}$
is associated with $R$-symmetry $U(1)_R$ \cite{Martelli:2004wu}.
Finally, the invariance with respect to the shift of the $\alpha$
angle corresponds to the $U(1)_F$ isometry \cite{Benvenuti:2004dy}.

The sets of coordinates $(\theta,\phi)$ and $(y,\beta)$ describe the base space $B_4$,
which is topologically the product $S^2 \times S^2$ \cite{Gauntlett:2004yd}.
The coordinate $\alpha$ then corresponds to an $S^1$ fibration over $B_4$
and the $5d$ space is topologically $S^2 \times S^3$ \cite{Gauntlett:2004yd}.
To construct a \emph{pp}-wave background we  will consider
a null geodesic lying on the poles of the two spheres. More precisely, 
we will put $\theta=0$ and $y=y_i$ with $i=1,2$.
This is analogous to the $T^{1,1}$ example, where the maximally 
supersymmeteric \emph{pp}-wave background also emerges in the Penrose limit
around a null geodesic located at the poles of the two spheres 
\cite{Itzhaki:2002kh,Gomis:2002km,PandoZayas:2002rx}.
In the $Y^{p,q}$ case there is only one $SU(2)$ and as
a consequence the BMN construction will be
different for $y=y_1$ and $y=y_2$.
We will use the following coordinate transformation:

\bea  \label{eq:CT1}
&&
t = \mu x^+ + \frac{x^-}{\mu R^2} \qquad
\rho = \frac{r}{R}      \qquad
y= y_i \left(1 - 3 \left(\frac{z_1}{R} \right)^2 \right)    \qquad
\theta = \left( \frac{6}{1 - c y_i} \right)^{\half} \frac{z_2}{R}
\nonumber \\
&&
\phi =   - \varphi_2  - \left(\mu x^+ - \frac{x^-}{\mu R^2}\right)   \qquad
\beta =  \frac{1}{y_i} \varphi_1 + c \varphi_2 + \left( c + \frac{1}{y_i} \right) 
                            \left(\mu x^+ - \frac{x^-}{\mu R^2}\right) 
\nonumber \\
&&
\psi^\prime = - \varphi_1 - \varphi_2 + \left(\mu x^+ - \frac{x^-}{\mu R^2}\right).
\eea
Plugging this into the $10$d metric and taking the limit $R \to \infty$
we get:

\be \label{eq:pp}
ds^2 = - 4 dx^+ dx^- + dr^2 + r^2 d \Omega_3^2 + dz_1^2 + z_1^2 d \varphi_1^2 + dz_2^2 + z_2^2 d \varphi_2^2
-  \mu ^2 \left( r^2 + z_1^2 + z_2^2 \right) d{x^+}^2.
\ee
Let us comment on the coordinate transformation (\ref{eq:CT1}).
The transformations of $t$ and $\rho$ in (\ref{eq:CT1})
are standard for backgrounds of the form $AdS_5 \times M_5$. Furthermore,  the transformations 
of $y$ and $\theta$ are well matched for a null geodesic lying at $y=y_i$ and $\theta=0$.
The unusual $R^{-2}$ scaling of $z_1$ can be understood by relating $y$ to the 
polar angle $\zeta$ of the sphere spanned by $y$ and $\psi$ \cite{Gauntlett:2004yd}:

\be
\cos \zeta(y) = \left( \frac{a-3y^2+2 c y^3}{a-y^2} \right)^{1/2}.
\ee
It is straightforward to see that expanding around $\zeta=\pi/2$
(which corresponds to $y=y_i$) we obtain regular $R^{-1}$ scaling for this coordinate.
Let us also comment on the connection to the conifold case.
The standard $T^{1,1}$ coordinates are related to the coordinates of (\ref{eq:5d})
by $\cos \theta_1 =y$, $\theta_2 =\theta$, $\phi_1=-\beta$ and $\phi_2=\phi$.
Substituting $c=0$, $a=3$ and $y_i=1$ (corresponding to $\theta_1=0$) into 
(\ref{eq:CT1}) we recover the transformation of \cite{Itzhaki:2002kh,Gomis:2002km,PandoZayas:2002rx}.

As was announced in the Introduction (\ref{eq:pp}) is the 
maximally supersymmetric \emph{pp}-wave background
which preserves all 32 supercharges. 
Exactly like in the conifold case the supersymmetry
is enhanced since the original geometry had only 4 supercharges in $10$d.
The background is also supported by a non-trivial RR 5-form:

\be
F_{(5)} = \mu \left( r dr \wedge d \Omega_3 + 
    z_1 dz_1 \wedge d \varphi_1 \wedge z_2 dz_2 \wedge d \varphi_2 \right) \wedge dx^+.
\ee

We will close this section by giving a relation between
the light-cone world-sheet Hamiltonian for the \emph{pp}-wave background (\ref{eq:pp}) 
and the currents associated with the isometries of the original background.
We have:

\be \label{eq:H1}
\frac{H}{\mu} = -\frac{p_+}{\mu} = i \partial_{x^+} = i \partial_t 
  - i \partial_\phi - \frac{i}{6} \left( 1 + \frac{c}{y_i} \right) \partial_\alpha 
  + i \left( \partial_\psi - \frac{c}{6} \partial_\alpha \right).
\ee

First in the global AdS coordinates we have $i \partial_t = \Delta$,
where $\Delta$ is the conformal dimension operator.
The derivative $J_3 \equiv -i \partial_\phi$ corresponds to the $T_3$-component 
of the $SU(2)$ current. Furthermore, we will denote the $U(1)_F$
charge $-i \ell \partial_\alpha$ by $J_\alpha$. 
Finally, the $R$-symmetry charge $2 i \partial_{\psi^\prime}$ is denoted by $J_R$. 
To summarize, we get:

\be \label{eq:H2}
\frac{H}{\mu} = \Delta - J
\qquad 
\textrm{where}
\qquad
J= - J_3 - \frac{1}{6 \ell} \left( 1 + \frac{1}{y_i}\right) J_\alpha + \half J_R
\ee
and we from now on we will set $\mu=1$.

%-----------------------------------------------------------------------------
\section{The field theory interpretation}
\label{FT}

Taking the Penrose limit corresponds to focusing on
chiral operators with large $\Delta$ and $J$ both scaling like
the \mbox{'t Hooft} coupling $\lambda = g_{YM}^2 N$, while keeping
the light-cone Hamiltonian $H= \Delta-J$ finite.
For a given $J$ there is a unique light-cone vacuum $H=0$. 
The corresponding operator in the dual gauge theory has the form $\textrm{Tr} \, \mathcal{O}^J$,
where trace is over the gauge indices.
The eight transverse $H=1$ excitations of the string are identified by
inserting $\Phi_{i=1,2,3,4}$ and $D_{a=1,2,3,4} \mathcal{O}$ into the trace 
\cite{Itzhaki:2002kh,Gomis:2002km,PandoZayas:2002rx}.
The goal of this section is to find the fields $\mathcal{O}$ and $\Phi_{i=1,2,3,4}$
for the case of the $Y^{p,q}$ field theory dual.

Before proceeding further, let us first briefly review the 
similar construction \cite{Itzhaki:2002kh,Gomis:2002km,PandoZayas:2002rx} 
for the Klebanov-Witten model \cite{Klebanov:1998hh}.
The gauge theory dual to the conifold
geometry is coupled to two
chiral bi-fundamental multiplets $(A_+,A_-)$ and $(B_+,B_-)$, which transform as a doublet of one of
the $SU(2)$'s each and are inert under the second $SU(2)$.
The conifold coordinates are related to these fields in the following way:

\be  \label{eq:uvxy}
u=A_+B_+, \quad v=A_-B_-, \quad  x=A_+B_-, \quad  y=A_-B_+
\ee
and the conifold definition $uv=xy$ appears as a consistency condition directly 
following from (\ref{eq:uvxy}).
The BMN operator $\textrm{Tr} \, (A_+B_+)^J$  was identified as the
dual to the light-cone Hamiltonian ground state $H=0$.
Moving the null geodesic from the north to the south poles of one of the $2$-spheres
amounts to replacing one of the fields $A_+$ or $B_+$ by $A_-$ or $B_-$ respectively. 
The first excited state $H=1$ of the
world-sheet Hamiltonian is degenerate and there are
eight BMN operators corresponding to this state. 
Six operators are  given by:

\be
D_{\mu=0,\ldots,3} \textrm{Tr} \, (A_+B_+)^J, \quad
 \textrm{Tr} \, A_+B_- (A_+B_+)^J \quad \textrm{and} \quad
 \textrm{Tr} \, A_-B_+ (A_+B_+)^J,
\ee
while the other two BMN operators are constructed by inserting the lowest
components of the two $SU(2)$ conserved currents into the $H=0$ operator $\textrm{Tr} \, (A_+B_+)^J$:

\be
 \textrm{Tr} A_+ \bar{A}_- (A_+B_+)^J \qquad \textrm{and} \qquad   \textrm{Tr} \bar{B}_+ B_- (A_+B_+)^J.
\ee
Although these operators are explicitly non-chiral they still have protected dimensions 
properly matching the $H=1$ condition as one can verify by exploring the KK
spectrum compactified on $T^{1,1}$ \cite{Gubser:2001zr,Itzhaki:2002kh}.

The quiver diagram of the gauge theory dual to the AdS${}_5 \times Y^{p,q}$ supergravity
background consists of nodes denoting $2 p$ gauge groups connected by $4p+2q$
arrows corresponding to various fields in bi-fundamental representations \cite{Benvenuti:2004dy}. 
There are six different types of fields:

\FIGURE[t]{
 \label{fig1}
\centerline{\begin{picture}(0,0)%
\includegraphics{nq1.pstex}%
\end{picture}%
\setlength{\unitlength}{2693sp}%
\begingroup\makeatletter\ifx\SetFigFont\undefined%
\gdef\SetFigFont#1#2#3#4#5{%
  \reset@font\fontsize{#1}{#2pt}%
  \fontfamily{#3}\fontseries{#4}\fontshape{#5}%
  \selectfont}%
\fi\endgroup%
\begin{picture}(5725,4522)(108,-3953)
\put(316,-1726){\makebox(0,0)[b]{\smash{{\SetFigFont{12}{14.4}{\rmdefault}{\mddefault}{\updefault}$6$}}}}
\put(4951,-736){\makebox(0,0)[b]{\smash{{\SetFigFont{12}{14.4}{\rmdefault}{\mddefault}{\updefault}$U_i$}}}}
\put(2926,-286){\makebox(0,0)[b]{\smash{{\SetFigFont{12}{14.4}{\rmdefault}{\mddefault}{\updefault}$Z$}}}}
\put(901,-736){\makebox(0,0)[b]{\smash{{\SetFigFont{12}{14.4}{\rmdefault}{\mddefault}{\updefault}$U_i$}}}}
\put(2926,-1411){\makebox(0,0)[b]{\smash{{\SetFigFont{12}{14.4}{\rmdefault}{\mddefault}{\updefault}$Y$}}}}
\put(4456,389){\makebox(0,0)[b]{\smash{{\SetFigFont{12}{14.4}{\rmdefault}{\mddefault}{\updefault}$2$}}}}
\put(1441,389){\makebox(0,0)[b]{\smash{{\SetFigFont{12}{14.4}{\rmdefault}{\mddefault}{\updefault}$1$}}}}
\put(901,-2761){\makebox(0,0)[b]{\smash{{\SetFigFont{12}{14.4}{\rmdefault}{\mddefault}{\updefault}$V_i$}}}}
\put(5626,-1681){\makebox(0,0)[b]{\smash{{\SetFigFont{12}{14.4}{\rmdefault}{\mddefault}{\updefault}$3$}}}}
\put(4951,-2761){\makebox(0,0)[b]{\smash{{\SetFigFont{12}{14.4}{\rmdefault}{\mddefault}{\updefault}$V_i$}}}}
\put(4501,-3751){\makebox(0,0)[b]{\smash{{\SetFigFont{12}{14.4}{\rmdefault}{\mddefault}{\updefault}$4$}}}}
\put(2926,-3886){\makebox(0,0)[b]{\smash{{\SetFigFont{12}{14.4}{\rmdefault}{\mddefault}{\updefault}$U_i$}}}}
\put(1396,-3841){\makebox(0,0)[b]{\smash{{\SetFigFont{12}{14.4}{\rmdefault}{\mddefault}{\updefault}$5$}}}}
\put(4051,-961){\makebox(0,0)[b]{\smash{{\SetFigFont{12}{14.4}{\rmdefault}{\mddefault}{\updefault}$Y$}}}}
\put(3601,-2311){\makebox(0,0)[b]{\smash{{\SetFigFont{12}{14.4}{\rmdefault}{\mddefault}{\updefault}$Y$}}}}
\put(2251,-2311){\makebox(0,0)[b]{\smash{{\SetFigFont{12}{14.4}{\rmdefault}{\mddefault}{\updefault}$Y$}}}}
\put(1801,-961){\makebox(0,0)[b]{\smash{{\SetFigFont{12}{14.4}{\rmdefault}{\mddefault}{\updefault}$Y$}}}}
\end{picture}%
}
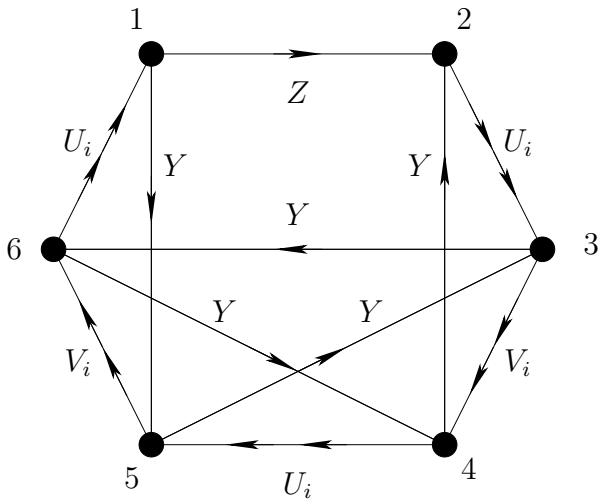
\caption{The quiver diagram of $Y^{3,2}$. 
} 
}

\FIGURE[b]{
 \label{fig2}
\centerline{\begin{picture}(0,0)%
\includegraphics{nq2.pstex}%
\end{picture}%
\setlength{\unitlength}{2171sp}%
\begingroup\makeatletter\ifx\SetFigFont\undefined%
\gdef\SetFigFont#1#2#3#4#5{%
  \reset@font\fontsize{#1}{#2pt}%
  \fontfamily{#3}\fontseries{#4}\fontshape{#5}%
  \selectfont}%
\fi\endgroup%
\begin{picture}(11872,1972)(365,-1497)
\end{picture}%
}
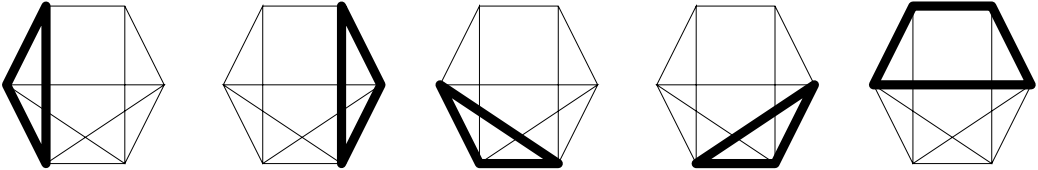
\caption{All $UVY$ and $UZUY$ ``short" operators of the $Y^{3,2}$ quiver theory.
The $F$-term conditions imply that all these operators are equivalent \cite{Benvenuti:2005cz}. 
} 
}

\begin{itemize}
\item $p$ $SU(2)$ doublets $U_{\alpha=1,2}$
\item $q$ $SU(2)$ doublets $V_{\alpha=1,2}$
\item $p+q$ singlets $Y$
\item $p-q$ singlets $Z$.
\end{itemize}
The quiver diagram for the special case of $p=3$ and $q=2$ is shown on Fig. \ref{fig1}.
The superpotential of the theory is built from various cubic and quartic "blocks"
that can be represented symbolically as $\textrm{Tr} \, UVY$ and $\textrm{Tr} \, UZUY$
respectively \cite{Benvenuti:2004dy}. 
In both cases the $SU(2)$ indices are contracted using the $\epsilon$-matrix.
The $F$-term relations derived from the superpotential
produce a set of non-trivial relations among the fields.
Using these relations one can construct the chiral ring of the gauge invariant 
operators \cite{Benvenuti:2005cz} (see also \cite{Berenstein:2005xa,Franco:2005zu,Bertolini:2005di}). 
In particular, each of the $p+q$ superpotential terms (both cubic and quartic)
has four gauge invariant operators naturally associated with it, namely
operators of the form $\textrm{Tr} \, U_\alpha V_\beta Y$ or $\textrm{Tr} \, U_\alpha ZU_\beta Y$
for $\alpha, \beta =1,2$.
The $F$-term conditions imply that all these operators are the same.
Moreover, the antisymmetric part of the $\half \bigotimes \half$ product identically vanishes.
Thus we end up with a single spin-1 gauge invariant ``short" operator $\mathcal{S}^{I=-1,0,-1}$
(see Fig. \ref{fig2}).  
The next chiral primary is obtained by multiplying all of the $U_\alpha$, $V_\alpha$ and $Z$ fields 
in clockwise direction along the quiver. This results in the so-called ``long"
operator  $\mathcal{L_+}$. Since the $F$-term conditions impose symmetrization over the 
$SU(2)$ indices, the only non-trivial component of this operator transforms in the 
$\half (p+q)$ representation of $SU(2)$. 
Finally, there is an additional ``long" operator  $\mathcal{L_-}$
built from the $U_\alpha$ and $Y$ fields. It transforms in the $\half (p-q)$ representation
(see Fig. \ref{fig3,4}).
Remarkably, the operators $\mathcal{L_+}$ and $\mathcal{L_-}$
have winding numbers $+1$ and $0$ with respect to the quiver diagram.

\FIGURE[b]{
 \label{fig3,4}
\centerline{\begin{picture}(0,0)%
\includegraphics{nq3.pstex}%
\end{picture}%
\setlength{\unitlength}{3355sp}%
\begingroup\makeatletter\ifx\SetFigFont\undefined%
\gdef\SetFigFont#1#2#3#4#5{%
  \reset@font\fontsize{#1}{#2pt}%
  \fontfamily{#3}\fontseries{#4}\fontshape{#5}%
  \selectfont}%
\fi\endgroup%
\begin{picture}(1972,1972)(590,-1722)
\end{picture}%
 \qquad \qquad \qquad \qquad \begin{picture}(0,0)%
\includegraphics{nq4.pstex}%
\end{picture}%
\setlength{\unitlength}{2368sp}%
\begingroup\makeatletter\ifx\SetFigFont\undefined%
\gdef\SetFigFont#1#2#3#4#5{%
  \reset@font\fontsize{#1}{#2pt}%
  \fontfamily{#3}\fontseries{#4}\fontshape{#5}%
  \selectfont}%
\fi\endgroup%
\begin{picture}(4447,1972)(590,-1722)
\end{picture}%
}
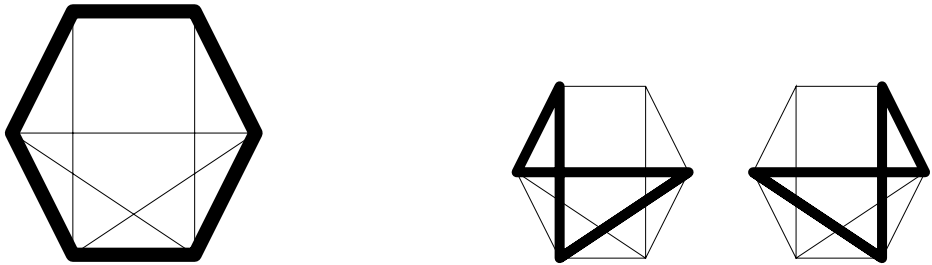
\caption{The $\mathcal{L_+}$ ``long" (left) and 
         the $\mathcal{L_-}$ ``long" (right) operator of $Y^{3,2}$. 
} 
}

The  charges of the operators are given by \cite{Benvenuti:2005cz}:

\be
\begin{tabular}{|c|c|c|c|}   
\hline   
    & spin $J$  & $J_R$ & $J_\alpha$ \\   
\hline    
\hline    
 $\mathcal{S}^I$  & 1  & 2 & 0 \\   
\hline    
 $\mathcal{L}_+ $  &  $\frac{p+q}{2}$  & $p+q-\frac{1}{3 \ell}$ & $1$ \\ 
\hline    
 $\mathcal{L}_- $  &  $\frac{p-q}{2}$  & $p-q+\frac{1}{3 \ell}$ & $-1$ \\ 
\hline\hline
\end{tabular}    
\ee

Substituting the charges of the lowest $SU(2)$ component ($J_3=-\half(p+q)$) of the 
$\mathcal{L_+}$  operator into the Hamiltonian (\ref{eq:H2}) we easily find that $H=0$ if $y=y_1$.
Therefore the string vacuum state in this case corresponds to the operator $\textrm{Tr} \, \mathcal{L}_+^J$. 
Analogously for $y=y_2$ the relevant operator is $\textrm{Tr} \, \mathcal{L}_-^J$.
In verifying this statement it is useful to recall that for the chiral primaries 
$\Delta = \frac{3}{2} \left \vert J_R \right \vert$
and:

\be   \label{eq:pqly}
p + q -\frac{1}{3 \ell} = -\frac{1}{3\ell} \frac{1}{y_1}
\qquad \textrm{and} \qquad
p - q +\frac{1}{3 \ell} = \frac{1}{3\ell} \frac{1}{y_2}.
\ee
Next let us consider the insertions corresponding to the eight $H=1$ states.
As we have mentioned above four insertion operators are given
by space derivatives $D_\mu \mathcal{L}_+$ and $D_\mu \mathcal{L}_-$
respectively. Therefore we have to identify four additional operators:

\begin{enumerate}
\item We can obtain $H=1$ by inserting the ``short" 
      operator $\mathcal{S}^{I=-1}$ as it follows immediately from the table (note that for $\mathcal{S}^{I=-1}$
      we have $J_3=-1$).
\item We took for the $H=0$ BMN operators the lowest components of the $\half(p+q)$ and $\half(p-q)$ $SU(2)$ representations 
      related to the $\mathcal{L}_+$ and $\mathcal{L}_-$ operators respectively. 
      It is natural therefore to consider a "spin flip" operator:
      we can change the spin of one of the doublets along the ``long"  operator then symmetrizing
      over all possible "flips". Since for the modified operator $\delta J_3=1$ with all other charges
      unchanged we find that it matches perfectly the $H=1$ condition.
      This is analogous to the $A_+ \to A_-$ and $B_+ \to B_-$ flips in the conifold case.
\item Consider the lowest component of the conserved $SU(2)$ current:

\be
\mathcal{K}^I_{SU(2)} = \sum_{i} \sigma_{\alpha \beta}^I \textrm{Tr} \left( U_\alpha^{i,i+1} \bar{U}_\beta^{i+1,i} + 
   V_\alpha^{i,i+1} \bar{V}_\beta^{i+1,i} \right),
\ee
    where $\sigma_{\alpha \beta}^{I=1,2,3}$ are the Pauli matrices.
	This operator has protected dimension $\Delta=2$ and vanishing $J_R$ and $J_\alpha$.
	On the other hand $J_3=-1,0,1$ and taking the lowest component ($J_3=-1$)
	as an insertion we find the required $H=1$ result. Again, as in the $T^{1,1}$ case,
	there is no apparent field theory argument protecting the naive counting and
	we have to analyze the supergravity spectrum with the given quantum numbers in order to
	verify the prediction. Unfortunately for an arbitrary $Y^{p,q}$ background it is quite difficult
	to carry out these calculations (for related discussions of the issue see \cite{Kihara:2005nt}
	and \cite{Oota:2005mr}). 

\item The last operator can be produced using the ``short-cut" operator \cite{Benvenuti:2005cz}. One starts
      from the lowest component of 
      ``long" operator $\mathcal{L}_+$, replaces a fragment $V_2^{i,i+1} U_2^{i+1,i+2}$, 
      $U_2^{i,i+1} V_2^{i+1,i+2}$ or $U_2^{i,i+1} Z U_2^{i+1,i+2}$ by the
      nearby antichiral $\bar{Y}_{i,i+2}$ field and finally symmetrizes by the "replacement"
      all over the quiver. For the new non-chiral operator we have $\delta J_\alpha=0$, $\delta J_R=-2$
      and $\delta J_3=-1$ (recall that we have replaced a symmetrized product of two $SU(2)$
      doublets by a singlet). Furthermore, we have $\delta \Delta=-1$. This might be expected
      from the fact that the
      ``short-cut" can be thought of as a combination of an insertion of the $U(1)_\alpha$ current 
      $\mathcal{K}_\alpha=\sum Y^{i,i+2} \bar{Y}^{i+2,i} + \ldots$
      and a removal of $\mathcal{S}^{I=-1}$. It is easy to "verify" that again $H=1$.
      Alternatively, for the case with the null geodesic at $y=y_2$ we can produce the $H=1$ operator by
      replacing one of the fragments
      $U_2^{i,i+1} Y^{i+1,i-1}$ of the ``long" operator 
      $\mathcal{L}^-$ (see Fig. \ref{fig3,4}) by the corresponding 
      anti-chiral field $\bar{V}^{i+2,i}$.
      Again, the complexity of the $Y^{p,q}$ background prevents us from 
      verifying this result by supergravity spectrum analysis and we refer the reader to the related
      papers \cite{Kihara:2005nt} and \cite{Oota:2005mr}.
\end{enumerate}
To summarize, the structure of the BMN operators of the $Y^{p,q}$ theories
is quite similar to the analogous construction in the Klebanov-Witten
$T^{1,1}$ model. Four out of eight operators corresponding to the $H=1$ excited state
are obtained by applying space-time derivatives on the ground state operator.
Two additional operators are produced by the spin "flip" and the $SU(2)$ current insertion
exactly like in the conifold example.
The last two operators (the ``short" operator insertion and the ``short-cut") differ, however,  
from the BMN construction in the KW model. This of course is a remnant of the
fact that there is only one $SU(2)$ factor in the symmetry group of the $Y^{p,q}$ model.

Notice that from four "non-derivative" operators two are chiral and the other two are
non-chiral precisely like in the conifold case.
In the rest of the section we will show that there is a straightforward way to identify
these chiral BMN operators using the fact that a Calabi-Yau cone over $Y^{p,q}$  is actually
a K\"ahler quotient $\mathbb{C}^4/\!\!/U(1)$, namely a gauged linear $\sigma$-model (GLSM) with
$U(1)$ charges $(p,p,-p+q,-p-q)$ \cite{Martelli:2004wu}. 
As one of the checks, exploring the chiral ring relations (as was briefly outlined above)
one arrives at the conclusion, that all the gauge-invariant chiral operators of the theory
are in one-to-one correspondence with $U(1)$-invariant polynomials of the GLSM.
These polynomials are of the form:

\be  \label{eq:pol}
P=w_1^{n_1} w_2^{n_2} w_3^{n_3} w_4^{n_4},
\ee
where $w_i$'s are the $\mathbb{C}^4$ coordinates and the non-negative integers $n_i$'s
satisfy the $U(1)$-invariance condition:

\be
p(n_1+n_2) - (p-q) n_3 - (p+q) n_4 = 0.
\ee
In particular, in the conifold case the coordinates $w_i$ may be identified directly
with the four fields $A_\pm$ and $B_\pm$. 
This just reflects the fact that  $F$-term conditions
derived from the Klebanov-Witten superpotential don't impose non-trivial relations between the fields.
For an arbitrary $Y^{p,q}$ there is certainly no
direct link between the fields $U_\alpha$, $V_\alpha$, $Y$, $Z$ 
and the $w_i$ coordinates of the corresponding GLSM,
and we can only map gauge invariant products of the fields
to the polynomials of the form (\ref{eq:pol}).
There are three types of independent polynomials for any $p$ and $q$:

\begin{enumerate}
\item $a_k = w_1^k w_2^{p-q-k} w_3^p$ with $k=0,\ldots,p-q$. This corresponds to the 
         $(p-q+1)$ components of the ``long" operator
        $\mathcal{L}_-$ discussed above.
\item $b_1 = w_1^2 w_3 w_4$, $b_2 = w_1 w_2 w_3 w_4$ and $b_3 = w_2^2 w_3 w_4$.
        These are the three components of the ``short" operator
        $\mathcal{S}^{\mathcal{I}=-1,0,1}$.
\item $c_k = w_1^k w_2^{p+q-k} w_4^p$ with $k=0,\ldots,p+q$. This corresponds to the ``long" operator
        $\mathcal{L}_+$.
\end{enumerate}
Next let us denote $\theta_i = \textrm{Arg}(w_i)$.
Note that $\partial_{\theta_i}=-i n_i$ while acting on the polynomials of the form (\ref{eq:pol}).
Moreover, the derivatives $\partial_{\theta_i}$ can be expressed in terms of the  
derivatives with respect to the angular coordinates appearing in the metric
of $Y^{p,q}$ (see \cite{Martelli:2004wu} for the detailed explanation):

\begin{eqnarray}   \label{eq:4Kv}
\partial_{\theta_1} &=& \partial_\phi + \partial_\psi      \nonumber \\
\partial_{\theta_2} &=& -\partial_\phi + \partial_\psi      \nonumber \\
\partial_{\theta_3} &=& \partial_\psi - \frac{\ell}{2 } (p+q) \partial_\alpha      \nonumber \\
\partial_{\theta_4} &=& \partial_\psi + \frac{\ell}{2 } (p-q) \partial_\alpha.    
\end{eqnarray} 
We can use these identities to express the derivatives $\partial_\phi$, $\partial_\psi$
and $\partial_\alpha$ in terms of the derivatives $\partial_{\theta_i}$'s.
Substituting further these relations into the expressions for $J_R$ and $J$ (see (\ref{eq:H2}))
we can re-write these currents solely in terms of the numbers $n_i=-i \partial_{\theta_i}$.
Finally, since for the chiral primaries operators $\Delta=\frac{3}{2} \left\vert J_R \right\vert$ we obtain 
the following simple identity for $H=\Delta-J$
(we will put $\mu=1$):

\be
H_{\textrm{C.P.}} = n_1+n_3 \quad \textrm{if} \quad y=y_1  \quad
\textrm{and} \quad
H_{\textrm{C.P.}} = n_1+n_4 \quad \textrm{if} \quad y=y_2.
\ee
Here the subscript ``C.P." reminds again that the relation is valid only for chiral primary operators
and we used (\ref{eq:pqly}) in the calculations.

We are now in a position to verify our results for the BMN operators dual to 
the $H=0$ and $H=1$ string states. 
For simplicity let us consider the $y=y_1$ case.
Since all $n_i$ in (\ref{eq:pol}) are non-negative,
$H_{\textrm{C.P.}}=0$ iff $n_1,n_3=0$. The only polynomial of this form is 
$c_0^N$ for arbitrary $N$. Since $c_0= w_2^{p+q} w_4^p$ is associated with the lowest
component of the ``long" operator $\mathcal{L}_+$, we successfully reproduce our result
for the BMN operator dual to the ground state. 
Furthermore, there are two options for the $H_{\textrm{C.P.}}=1$ state.
Namely, for $n_1=1$, $n_3=0$ the corresponding polynomial is $c_1=w_1 w_2^{p+q-1} w_4^p$
and for $n_1=0$, $n_3=1$ the polynomial is $b_3=w^2_2 w_3 w_4 $. Since the former corresponds 
to the spin flip and the latter to the lowest component of the ``short" operator $\mathcal{S}^{I=-1}$ we
recover the above-mentioned result for the two chiral $H=1$ states.  
Finally, let us address the ``short-cut" $H=1$ non-chiral operator. As we have just discussed this operator is
obtained by the current $\mathcal{K}_\alpha$ insertion into the $\textrm{Tr} \, \mathcal{L}_+^J$
string followed by the $\mathcal{S}^{I=-1}$ operator removal. From the current insertion we get
$\delta H=2$, so we need $\delta H=1$ for the chiral operator $\mathcal{S}^{I=-1}$. From
the discussion above this is clearly the case, since $\mathcal{S}^{I=-1}$ corresponds
to the $b_3=w_2^2 w_3 w_4$ polynomial with $n_1 +n_3=1$. 
The reader may wonder if it is possible to construct another ``short-cut" $H=1$ operator 
starting from the same current, but removing another chiral operator from the string (for instance, $S^{I=+1}$).
An easy check reveals, however, that the $S^{I=-1}$ ``short-cut" is the only possibility.

%-----------------------------------------------------------------------------
\section{The $L^{p,q,r}$ spaces case}
\label{Lpqr}

In this section we will apply the method proposed above to the $L^{p,q,r}$ case.
This is a larger family of backgrounds 
with only $U(1)^3$ isometry group, which include the $Y^{p,q}$ sub-family as a special case.
For general $p,q$ and $r$ the gauge theory field content is extremely
complicated and we will not try to present it here 
(see \cite{Benvenuti:2005ja,Butti:2005sw,Franco:2005sm}).
Instead, we will use the underlying K\"ahler quotient structure of 
the space exactly as we did for $Y^{p,q}$ in the previous section.
First we will show that the Penrose limit again provides the \emph{pp}-wave
metric (\ref{eq:pp}). Then we will re-write the light-cone Hamiltonian 
in terms of the derivatives (\ref{eq:4Kv}) and will use this presentation to
identify the BMN operator dual to the ground state and 
two chiral "non-derivative" BMN operators corresponding to the first excited state.

We will start with a very brief review of the $L^{p,q,r}$ geometry.
The relevant $5$d Sasaki-Einstein metric is \cite{Cvetic:2005ft,Martelli:2005wy}:

\bea
\frac{1}{R^2} ds^2_{L^{p,q,r}} &=& (d \tau + \sigma)^2 + \frac{\rho^2 dx^2}{4 \Delta_x}
                                +  \frac{\rho^2 d\theta^2}{ \Delta_\theta}       
                                + \frac{\Delta_x}{\rho^2} \left(
                                     \frac{\sin ^2 \theta}{\alpha} d\phi + \frac{\cos ^2 \theta}{\beta} d\psi
														\right)^2   +                \nonumber \\
&&
		+ 	\frac{\Delta_\theta \sin^2 \theta \cos^2 \theta}{\rho^2} \left(
                                     \frac{\alpha-x}{\alpha} d\phi - \frac{\beta-x}{\beta} d\psi
														\right)^2	,																							
\eea
where 

\bea
\sigma &=& \frac{(\alpha-x) \sin ^2 \theta}{\alpha} d\phi + \frac{(\beta-x) \cos ^2 \theta}{\beta} d\psi,
  \nonumber \\
\Delta_x &=& x(\alpha-x)(\beta-x)-1,  \qquad \Delta_\theta= \alpha \cos^2 \theta + \beta \sin^2 \theta 
\quad \textrm{and} 
\quad \rho^2 = \Delta_\theta-x.
\eea
The constants $\alpha$ and $\beta$ as well as the period of the angular coordinate 
$\tau$ are very complicated functions of the three co-prime integer parameters
$p$, $q$ and $r$ \cite{Cvetic:2005vk,Martelli:2005wy} satisfying $p < r < q$. 
In particular, for $p+q=2r$ one has $\alpha=\beta$ and the geometry reduces 
to the $Y^{p^\prime,q^\prime}$ case with $p^\prime=\frac{1}{2}(p+q)$ and $q^\prime=\frac{1}{2}(q-p)$.
The $x$ coordinate ranges between $x_1$ and $x_2$, the lowest two roots of the equation $\Delta_x=0$.
Moreover, $0 \le \theta \le \frac{\pi}{2}$.

The $4$-dimensional base of the $L^{p,q,r}$ space is topologically the product $S^2 \times S^2$
exactly like in the $Y^{p,q}$ case.
The coordinates $x$ and $\theta$ are the azimuthal coordinates on the two $2$-spheres.
We will again assume that the null geodesic is located at the poles of the spheres,
namely $x=x_1$ or $x_2$ and $\theta=0$ or $\frac{\pi}{2}$ along the geodesic.
Since there is no $SU(2)$ isometry like in the $Y^{p,q}$ geometry, taking the Penrose limit
might yield four different interpretations on the field theory side depending on the four possible locations 
of the geodesic.
For the sake of simplicity in what follows we will only consider the $\theta=0$ option. 

The coordinate transformation we will use is:

\bea  \label{eq:CT2}
&&
x= x_i + \frac{\Delta^\prime_i}{\alpha-x_i} \frac{z_1^2}{R^2}     \qquad  \qquad
\theta = \left( \frac{\alpha}{\alpha-x_i} \right)^{\half} \frac{z_2}{R}
\nonumber \\
&&
\phi =  a_i \varphi_1 + \varphi_2 +  \left( 1 + a_i \right)
                                        \left(\mu x^+ - \frac{x^-}{\mu R^2}\right)  
\nonumber \\
&&                                        
\psi =  b_i \varphi_1 + b_i \left(\mu x^+ - \frac{x^-}{\mu R^2}\right)  
\nonumber \\
&&
\tau = c_i \varphi_1 + \left( 1 + c_i \right)\left(\mu x^+ - \frac{x^-}{\mu R^2}\right),  
\eea
where $i=1$ or $2$
and the
transformation of the $AdS_5$ coordinates is the same as in the $Y^{p,q}$ case. 
The constants $a_i$, $b_i$ and $c_i$ are given by:

\be
a_i = \frac{\alpha(\beta-x_i)}{\Delta^\prime_i}, \quad
b_i = \frac{\beta(\alpha-x_i)}{\Delta^\prime_i},  \quad
c_i = - \frac{(\beta-x_i)(\alpha-x_i)}{\Delta^\prime_i}
\quad \textrm{with} \quad 
\Delta^\prime_i \equiv \left. \frac{\partial \Delta_x}{\partial x} \right \vert_{x=x_i}.
\ee
From \ref{eq:CT2} we find that $\partial_{\varphi_1}=l_i$, where 
$l_i=a_i \partial_{\phi} + b_i \partial_{\psi} + c_i \partial_{\tau}$ is the Killing vector 
whose length vanishes at $x=x_i$ \cite{Cvetic:2005ft}. Similarly  $\partial_{\varphi_1}=\partial_\phi$ is the
Killing vector whose norm equal to zero at $\theta=0$. It means that if all triples among 
the integers $(p,q,r,s)$ are co-prime, and as a consequence there will be no conical singularities,
than the periods of both $\varphi_1$ and $\varphi_2$ will be $2 \pi$. 

Substituting (\ref{eq:CT2}) into the $10$d metric and taking the $R \to \infty$
limit yields the maximally supersymmetric \emph{pp}-wave background (\ref{eq:pp})
and the light-cone world-sheet Hamiltonian is given by:

\be   \label{eq:H3}
\frac{H}{\mu} = i \partial_{x^+} = \Delta - J,
\ee
where $\Delta= i \partial_t$ as usual and 
\be
J = -i b_i \partial_\psi - i \left( 1 + a_i \right) \partial_\phi
      -i \left( 1 + c_i \right) \partial_\tau.                                       
\ee

Now we will use the method described in the previous section in
order to identify the chiral BMN operators dual to the ground
and the first excited states of the light-cone Hamiltonian (\ref{eq:H3}).
A Calabi-Yau cone over the $L^{p,q,r}$ base is a K\"ahler quotient
with charges $(p,q,-r,r-p-q)$ and the set of four Killing vectors analogous to the set
(\ref{eq:4Kv}) is given by \cite{Cvetic:2005vk}:

\bea
\partial_{\theta_i} &=&  - \left( c_i \partial_\tau + a_i \partial_\phi + b_i \partial_\psi \right)
\qquad \textrm{for} \quad i=1, 2      
 \nonumber \\
\partial_{\theta_3} &=& -\partial_\phi    \nonumber \\
\partial_{\theta_4} &=& -\partial_\psi.
\eea
Furthermore, the $R$-charge current is:

\be
J_R = - \frac{2}{3} i \partial_\tau.
\ee
Plugging these identities into (\ref{eq:H3}) and recalling again that
for chiral primary operators $\Delta = \frac{3}{2} \left \vert J_R \right \vert$
we arrive at the following result for $\mu=1$:

\be  \label{eq:HbpsLpqr}
H_{\textrm{C.P.}} = n_1+n_3 \quad \textrm{if} \quad x=x_1  \quad
\textrm{and} \quad
H_{\textrm{C.P.}} = n_2+n_3 \quad \textrm{if} \quad x=x_2.
\ee
Notice that for a null geodesic lying at $\theta=\frac{\pi}{2}$
we would get the same result with $n_3$ replaced by $n_4$.

It is now a straightforward exercise to find the polynomials 
of $w_i$'s, which will correspond to $H=0$ and $H=1$.
Let us focus on the $x=x_1$ case. As in the $Y^{p,q}$ case
the condition $H_{\textrm{C.P.}}=0$ implies that $n_1=n_3=0$ and
hence the relevant polynomial is $w_2^s w_4^q$ with $s=p+q-r$.
Finally, for $H_{\textrm{C.P.}}=1$ we have polynomials corresponding to $(n_1,n_3)=(1,0)$
and $(n_1,n_3)=(0,1)$. The first polynomial is $w_1 w_2^a w_4^b$ with
$p+qa-sb=0$ and the second is $w_3 w_2^c w_4^d$ with
$r+qc-sd=0$. Notice that in both cases the solutions of the Euclidean equations exist, 
since the integers $q$ and $s$ are co-prime.
As we explained above, for not co-prime $q$ and $s$
the period of the  angular coordinates $\varphi_1$ and  $\varphi_2$
will be different from to $2 \pi$ changing the string spectrum in the \emph{pp}-wave background
(\ref{eq:pp})\footnote{
In this case the calculation of the spectrum will be similar to the 
Penrose limit of the orbifold of $AdS_5 \times S^5$
(see \cite{Mukhi:2002ck,Kuperstein:2003jd} and also \cite{Sadri:2005gi}.)}.

Next let us address the remaining two non-chiral operators corresponding to the $H=1$ state.
Unlike in the $Y^{p,q}$ case here we do not have $SU(2)$ symmetry and we therefore cannot use the 
related current as an insertion in order to construct the relevant $H=1$ operator.
On the other hand, we have two independent $U(1)$ currents and so we might attempt to
build two appropriate ``short-cut" non-chiral operators for each one of the currents.
Exactly like in the $Y^{p,q}$ case we have $\delta H=2$ for these currents, since they are
invariant under the $U(1)$ isometries of the theory, while $\Delta=2$ by the field theory arguments.
It implies again that we are looking for two chiral operators with 
$\delta H_\textrm{C.P.}=1$ or $n_1+n_3=1$. These are actually precisely the $H=1$ 
operators corresponding to the polynomials $(n_1,n_3)=(1,0)$ and $(n_1,n_3)=(0,1)$ 
that we have discussed in the previous paragraph. This statement, however, still needs to be  
verified by an explicit construction in terms of the field theory operators similarly to what we did
in the $Y^{p,q}$ case.

%-----------------------------------------------------------------------------
\section{The $L^{1,7,3}$ space example}
\label{L1735}

Unfortunately, we were not able to
construct explicitly the BMN operators dual to the ground and the first excited states
for arbitrary parameters $(p,q,r,s)$.
Instead we will thoroughly analyze the $L^{1,7,3}$ example ($p=1,q=7,r=3$ and $s=5$).
To this end we have to identify the polynomials of the GLSM with the charges $(1,7,-3,-5)$
in terms of the gauge invariant field theory operators.  

As in the previous sections we will denote the K\"ahler quotient coordinates by
$w_1$, $w_2$, $w_3$ and $w_4$ with the charges $1$, $7$, $-3$ and $-5$ respectively.
A simple straightforward calculation shows that for these charges 
one has 12 independent polynomials\footnote{
By independence of the polynomials we mean that none of them can be written in terms 
of other polynomials from the list, namely there is no relation of the form $p=p_1^{n_1}p_2^{n_2}$ \ldots } :

\bea    \label{eq:www}
&& 
a = w_1^3 w_3, \qquad  b = w_2^5 w_4^7, \qquad c = w_1^5 w_4, \qquad  d = w_2^3 w_3^7,
\qquad e = w_1 w_2 w_3 w_4
\nonumber \\
&& f_1 = w_1 w_2^2 w_3^5, \qquad f_2 = w_1 w_2^2 w_4^3, \qquad f_3 = w_1^2 w_2 w_3^3, \qquad f_4 = w_1^3 w_2 w_4^2, 
\nonumber \\
&& \qquad f_5 = w_2^4 w_3 w_4^5, \qquad f_6 = w_2^2 w_3^3 w_4, \qquad f_7 = w_2^3 w_3^2 w_4^3.
\eea

Since the complex space described by the variables is only $3$-dimensional, these variables
are subject to many redundant relations between them. Here we will list only a few of them:

\be   
a^{35} b^3 = c^{21} d^5,   \label{eq:relations1} 
\ee
\be
e^8 = a b c d,   \label{eq:relations2} 
\ee
\be
f_1^3= ad^2 ,\quad
f_2^5= cb^2 ,\quad
f_3^3= a^2d ,\quad
f_4^5= c^3b ,\quad
f_5^7= b^5d ,\quad
f_6^7= bd^3 ,\quad
f_7^7= b^3d^2.       \label{eq:relations3}
\ee
The relations (\ref{eq:relations1}) and (\ref{eq:relations2}) can be easily generalized to arbitrary
parameters $p$, $q$ $r$ and $s=p+q-r$. Indeed, defining:

\be
a = w_1^r w_3^p, \qquad  b = w_2^s w_4^q, \qquad c = w_1^s w_4^p, \qquad  d = w_2^r w_3^q,
\qquad e = w_1 w_2 w_3 w_4
\ee
we get:

\be \label{eq:relationsGEN}
a^{qs} b^{pr} = c^{qr} d^{ps} \qquad \textrm{and} \qquad e^{p+q} = abcd.
\ee
Let us next briefly review the field theory content and the superpotential of the gauge theory dual
of an arbitrary $L^{p,q,r}$ background.
There are six types of fields that we
will  denote\footnote{In this paper we will follow mostly the notations of \cite{Franco:2005sm} 
exchanging only the fields $V_1$ and $V_2$ in order to make explicit the similarity to the $Y^{p,q}$ case.}
by $U_1$, $U_2$, $V_1$, $V_2$, $Y$ and $Z$ following the $Y^{p,q}$ conventions.  
The multiplicities of these fields are 
given by:

\FIGURE[t]{
 \label{L1753}
\centerline{\begin{picture}(0,0)%
\includegraphics{l1753.pstex}%
\end{picture}%
\setlength{\unitlength}{3947sp}%
\begingroup\makeatletter\ifx\SetFigFont\undefined%
\gdef\SetFigFont#1#2#3#4#5{%
  \reset@font\fontsize{#1}{#2pt}%
  \fontfamily{#3}\fontseries{#4}\fontshape{#5}%
  \selectfont}%
\fi\endgroup%
\begin{picture}(5007,4313)(10,-4011)
\put(2489,-61){\makebox(0,0)[b]{\smash{{\SetFigFont{14}{16.8}{\rmdefault}{\mddefault}{\updefault}$Z$}}}}
\put(4376,-1986){\makebox(0,0)[lb]{\smash{{\SetFigFont{14}{16.8}{\rmdefault}{\mddefault}{\updefault}$U_1$}}}}
\put(3626,-961){\makebox(0,0)[lb]{\smash{{\SetFigFont{14}{16.8}{\rmdefault}{\mddefault}{\updefault}$U_1$}}}}
\put(3751,-3211){\makebox(0,0)[lb]{\smash{{\SetFigFont{14}{16.8}{\rmdefault}{\mddefault}{\updefault}$Y$}}}}
\put(2426,-3911){\makebox(0,0)[b]{\smash{{\SetFigFont{14}{16.8}{\rmdefault}{\mddefault}{\updefault}$U_1$}}}}
\put(1201,-3186){\makebox(0,0)[rb]{\smash{{\SetFigFont{14}{16.8}{\rmdefault}{\mddefault}{\updefault}$Y$}}}}
\put(651,-1986){\makebox(0,0)[rb]{\smash{{\SetFigFont{14}{16.8}{\rmdefault}{\mddefault}{\updefault}$U_1$}}}}
\put(1414,-936){\makebox(0,0)[rb]{\smash{{\SetFigFont{14}{16.8}{\rmdefault}{\mddefault}{\updefault}$U_1$}}}}
\put(3314,-636){\makebox(0,0)[lb]{\smash{{\SetFigFont{14}{16.8}{\rmdefault}{\mddefault}{\updefault}$Y$}}}}
\put(1664,-636){\makebox(0,0)[rb]{\smash{{\SetFigFont{14}{16.8}{\rmdefault}{\mddefault}{\updefault}$Y$}}}}
\put(2339,-2336){\makebox(0,0)[rb]{\smash{{\SetFigFont{14}{16.8}{\rmdefault}{\mddefault}{\updefault}$U_2$}}}}
\put(2614,-2361){\makebox(0,0)[lb]{\smash{{\SetFigFont{14}{16.8}{\rmdefault}{\mddefault}{\updefault}$U_2$}}}}
\put(1614,-1611){\makebox(0,0)[lb]{\smash{{\SetFigFont{14}{16.8}{\rmdefault}{\mddefault}{\updefault}$V_2$}}}}
\put(3414,-1611){\makebox(0,0)[rb]{\smash{{\SetFigFont{14}{16.8}{\rmdefault}{\mddefault}{\updefault}$V_2$}}}}
\put(2476,-1324){\makebox(0,0)[b]{\smash{{\SetFigFont{14}{16.8}{\rmdefault}{\mddefault}{\updefault}$Y$}}}}
\put(2476,-2624){\makebox(0,0)[b]{\smash{{\SetFigFont{14}{16.8}{\rmdefault}{\mddefault}{\updefault}$U_2$}}}}
\put(3601,-1974){\makebox(0,0)[b]{\smash{{\SetFigFont{14}{16.8}{\rmdefault}{\mddefault}{\updefault}$Y$}}}}
\put(1426,-1999){\makebox(0,0)[b]{\smash{{\SetFigFont{14}{16.8}{\rmdefault}{\mddefault}{\updefault}$Y$}}}}
\put(3226,-3086){\makebox(0,0)[rb]{\smash{{\SetFigFont{14}{16.8}{\rmdefault}{\mddefault}{\updefault}$V_1$}}}}
\put(1676,-3011){\makebox(0,0)[lb]{\smash{{\SetFigFont{14}{16.8}{\rmdefault}{\mddefault}{\updefault}$V_1$}}}}
\put(1701,-3336){\makebox(0,0)[b]{\smash{{\SetFigFont{14}{16.8}{\rmdefault}{\mddefault}{\updefault}$V_2$}}}}
\put(3226,-3361){\makebox(0,0)[b]{\smash{{\SetFigFont{14}{16.8}{\rmdefault}{\mddefault}{\updefault}$V_2$}}}}
\end{picture}%
}
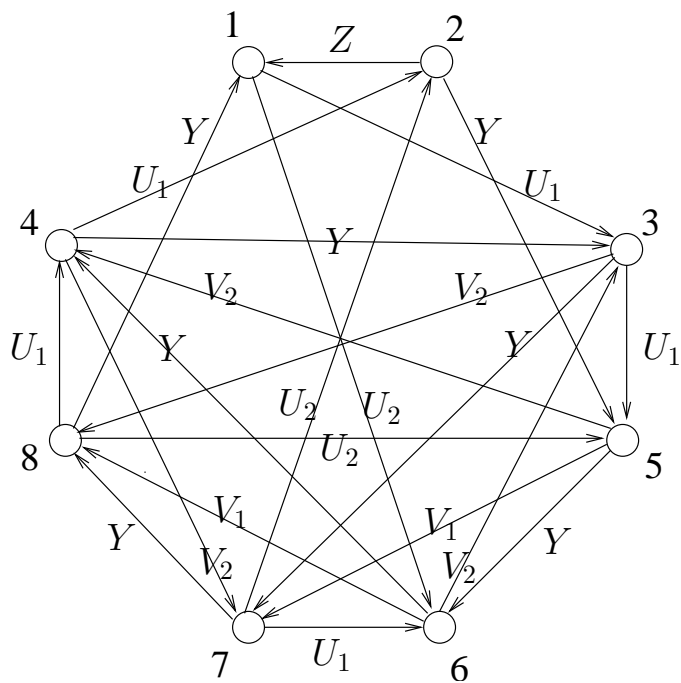
\caption{The quiver diagram of $L^{1,7,3}$ ($p=1,q=7,r=3$ and $s=5$). 
} 
}

\bea
\textrm{mult}[U_1] = s,    \quad \textrm{mult}[U_2] = r,   \quad  &&
\textrm{mult}[V_1] = r-p,  \quad \textrm{mult}[V_2] = q-r, \quad    
\nonumber \\
\textrm{mult}[Y] = q, && \quad \textrm{mult}[Z] = p. 
\eea
In particular, for $r=s$ the multiplicities of $U_1$ and $U_2$ become equal reproducing correctly
the number of the $SU(2)$ doublets $(U_1,U_2)$ in the $Y^{p^\prime,q^\prime}$ theory with 
$p^\prime=\half(p+q)=r$. Similar checks can be easily performed for the fields $V_1$ and $V_2$.
The quiver diagram of $L^{1,7,3}$ is depicted in Fig. \ref{L1753}.
Furthermore, exactly like in the $Y^{p,q}$ case there are three types of 
polynomials constructed from these fields that may appear in the superpotential:

\be  \label{eq:blocks}
W_0 = \textrm{Tr} \, Y U_1 Z U_2 , \qquad W_1 = \textrm{Tr} \, Y U_1 V_2 \quad 
\textrm{and} \quad W_2 = \textrm{Tr} \, Y U_2 V_1.
\ee
The number each term appears in the superpotential are $2p$, $2(q-r)$ and $2(r-p)$
respectively. The explicit form of the superpotential is quite complicated, but it can be figured out
in a straightforward manner using the corresponding dimer tiling (see \cite{Franco:2005sm}).
For the given $L^{1,7,3}$ example, however, the superpotential blocks (\ref{eq:blocks}) can be read
directly from the quiver diagram on Fig. \ref{L1753}. 

Differentiating the superpotential with respect to the fields we will obtain a set of $F$-term conditions.
Unlike in the conifold example these conditions are non-trivial and impose relations between various 
operators constructed from the fields. The task of constructing all possible gauge invariant field 
polynomials (operators) is very cumbersome
already for the $L^{1,7,3}$ case, and here we will report
only the final results.  

First, there are four polynomials analogous to the ``long" operators $\mathcal{L}_{\pm}$ which appeared 
in the $Y^{p,q}$ diagrams (see Fig. \ref{fig3,4}). They can be written in a schematic way as:

\bea   \label{eq:L+-}
\mathcal{L}^\uparrow_+ = \textrm{Tr} \, Z^p \left( U_1 \right)^s \left( V_1 \right)^{q-s}, &\quad&
\mathcal{L}^\downarrow_+ =\textrm{Tr} \, Z^p \left( U_2 \right)^r \left( V_2 \right)^{q-r}, 
\nonumber \\
\mathcal{L}^\uparrow_- =\textrm{Tr} \,  \left( U_1 \right)^p Y^r, &\quad&
\mathcal{L}^\downarrow_- =\textrm{Tr} \,  \left( U_2 \right)^p Y^s.
\eea
Here the arrows indicate that for $r=s$ the operators $\mathcal{L}^\uparrow_\pm$ reduce to the 
highest $SU(2)$ components of the ``long" operators $\mathcal{L}_\pm$ of the related $Y^{p^\prime,q^\prime}$
theory. Similarly, the operators $\mathcal{L}^\downarrow_\pm$ become the lowest components of
$\mathcal{L}_\pm$.
Using the diagram in Fig. \ref{L1753} it is quite easy to construct the $\mathcal{L}^\uparrow_+$
and $\mathcal{L}^\downarrow_+$ operators explicitly for the $L^{1,7,3}$ case:

\be
\mathcal{L}^\uparrow_+   = \textrm{Tr} \, Z_{21} U_{16}^2 V^2_{63} V^2_{38}  U_{85}^2 V_{54}^2 V_{47}^2 U_{72}^2, \quad
\mathcal{L}^\downarrow_+ = \textrm{Tr} \, Z_{21} U_{13}^1 U^1_{35} V_{57}^1 U_{76}^1 V_{68}^1 U_{84}^1 U_{42}^1.
\ee
Like in the $Y^{p,q}$ case these two operators have single representations in terms of the fields $U_i$,$V_i$,
$Y$ and $Z$.
The operators $\mathcal{L}^\uparrow_-$ and $\mathcal{L}^\downarrow_-$, however,  
have many possible representations, again similar to the $Y^{p,q}$ example. For instance,
the operator $\mathcal{L}^\uparrow_-$ may be written in five equivalent ways as can be 
shown by analyzing the set of $F$-term conditions:

\bea  \label{eq:Lup-}
\mathcal{L}^\uparrow_- = && \textrm{Tr} \, U_{84}^1 Y_{43} Y_{37} Y_{78}, \quad
                            \textrm{Tr} \, U_{13}^1 Y_{37} Y_{78} Y_{81}, \quad
                            \textrm{Tr} \, U_{42}^1 Y_{25} Y_{56} Y_{64},
\nonumber \\                         
                         && \qquad \textrm{Tr} \, U_{76}^1 Y_{64} Y_{43} Y_{37}, \quad
                                   \textrm{Tr} \, U_{35}^1 Y_{56} Y_{64} Y_{43}.
\eea
Using the $F$-term relations it is quite easy to show that these four operators 
are equivalent to each other. For example, in order to prove the equivalence of the first two operators
in (\ref{eq:Lup-})
it is enough to replace $U_{84}^1 Y_{43}$ by $Y_{81} U^1_{13}$ 
using the $F$-term condition for the field $V^2_{38}$.
The operator $\mathcal{L}^\downarrow_-$ also can be presented in various ways:

\be
\mathcal{L}^\downarrow_- = \textrm{Tr} \, U_{85}^2 Y_{56} Y_{64} Y_{43} Y_{37} Y_{78}, \quad
                             \textrm{Tr} \, U_{16}^2 Y_{64} Y_{43} Y_{37} Y_{78} Y_{81}, \quad
                             \textrm{Tr} \, U_{72}^2 Y_{25} Y_{56} Y_{64} Y_{43} Y_{37}.
\ee 
The operators $\mathcal{L}^\uparrow_\pm$ and $\mathcal{L}^\downarrow_\pm$
were called extremal BPS mesons in \cite{Benvenuti:2005ja}. Indeed, it can be verified
that these operators have maximal $U(1)$ charges (in modulus) for given $R$-charge.
Furthermore, it was argued in \cite{Benvenuti:2005ja} that these four mesons correspond
to the BPS geodesics, which lie at the vertices of the coordinate rectangular, namely at
$x=x_1$ or $x_2$ and $\theta=0$ or $\frac{\pi}{2}$. In terms of the K\"ahler quotient
coordinates the vertices are given by $w_i=w_j=0$, where $i=1$ or $2$ and $j=3$ or $4$. 
Therefore these vertices are well described by the polynomials $a$, $b$, $c$ and $d$ from
(\ref{eq:www}). For example, for $w_2=w_4=0$ the only non-vanishing polynomial in 
(\ref{eq:www}) is $a$. Thus we find that it is natural to relate the extremal BPS mesons  
 $\mathcal{L}^\uparrow_\pm$ and $\mathcal{L}^\downarrow_\pm$ to the variables $a$, $b$, $c$ and $d$.
To be more precise the identification is as follows:

\be  \label{eq:identification}
\mathcal{L}^\uparrow_+ \longleftrightarrow  d, \qquad 
\mathcal{L}^\downarrow_+ \longleftrightarrow  b, \qquad
\mathcal{L}^\uparrow_- \longleftrightarrow  a, \qquad
\mathcal{L}^\downarrow_- \longleftrightarrow  c.
\ee
To prove this statement one can verify, for instance, that $R$-charges of the
corresponding operators and polynomials in (\ref{eq:identification}) coincide with each other. 
Alternatively we can check (\ref{eq:identification}) by 
substituting the operators $\mathcal{L}^{\uparrow \downarrow}_\pm$ 
instead of $a$, $b$, $c$ and $d$ into the relation (\ref{eq:relations1}) and proving it using
the $F$-term conditions.
We will not perform this tedious calculation here.
Instead we will confirm  (\ref{eq:relations1}) by examining the quantum numbers of the operators.
Let us assign the following quantum numbers to the gauge theory fields:

\be
\label{eq:charges}
\begin{tabular}{|c|c|c|c|c|c|c|} 
\hline   
    & $U_1$ & $U_2$ & $V_1$ & $V_2$ & $Z$ & $Y$ \\   
\hline    
\hline    
 $Q_1$   & $\half$  & $-\half$ & $\half$ & $-\half$ & $0$       & $0$     \\ 
\hline 
 $Q_2$   & $1$      & $1$      & $1$     & $1$      & $0$       & $0$      \\   
\hline       
 $Q_3$   & $0$      & $0$      & $0$     & $0$      & $0$       & $1$      \\ 
\hline\hline
\end{tabular}    
\ee

Clearly, the superpotential is invariant under these $U(1)$ symmetries, since all of
the superpotential blocks in (\ref{eq:blocks}) have the same charges $(0,2,1)$ with respect
to (\ref{eq:charges}).  These $U(1)$'s are actually linear combinations of the $U(1)_R$, $U(1)_B$
and the other two $U(1)$ global symmetries of \cite{Benvenuti:2005ja,Franco:2005sm,Butti:2005sw},
but for what follows we will not need any relation between the symmetries of \cite{Benvenuti:2005ja,Franco:2005sm,Butti:2005sw}
and the charges of the table (\ref{eq:charges}). Substituting these charges into  
(\ref{eq:L+-}) we will get the charges of the ``long" operators $\mathcal{L}_\pm^\uparrow$
and $\mathcal{L}_\pm^\downarrow$. It is now a simple exercise to confirm that
with the identification (\ref{eq:identification})
the left and the right hand sides of  (\ref{eq:relations1}) 
(or  (\ref{eq:relationsGEN}) for arbitrary $p$, $q$ and $r$)  have the same
charges. This provides a very non-trivial check of (\ref{eq:identification}).

We can use the same method in order to find operators corresponding to the 
variables $e$ and $f_{i=1...7}$. Indeed, we see from (\ref{eq:relations2}) that
the $U(1)$ numbers (\ref{eq:charges}) of the operators corresponding to $e$ are $(0,2,1)$. These operators,
therefore, are just the superpotential blocks (\ref{eq:blocks}). There
are $2q$ blocks in general and their equivalence almost trivially follows from the
$F$-term conditions. Remarkably, there are precisely two blocks for each $Y_{ij}$ field.
For instance, for $Y_{43}$ we have $\textrm{Tr} \, Y_{43} U^1_{35} V^2_{54}$  
and $\textrm{Tr} \, Y_{43} V^2_{38} U^1_{84}$. Let us next consider the variable $f_7$.
From (\ref{eq:charges}) and the last equation in (\ref{eq:relations2}) we see that the charges of 
the corresponding operator are $(-\half,5,0)$. We found two gauge invariant products of the fields
with these quantum numbers:

\be   \label{eq:f7}
f_7 \qquad \longleftrightarrow \qquad \textrm{Tr} \, U^1_{42} Z_{21} U_{13}^1 V^2_{38} U_{85}^2 V_{54}^1 
\quad \textrm{and} \quad \textrm{Tr} \, U^2_{72} Z_{21} U_{13}^1 V^2_{38} U_{84}^1 V_{47}^2.
\ee
Let us show as a simple exercise that if one imposes
the $F$-term conditions, the two operators above become equivalent. Indeed, from the $F$-term condition for the field $Y_{25}$ it is clear that
we can replace the $V^2_{54} U^1_{42}$ in the first sequence of the fields by
$V^1_{57} U^2_{72}$. Next, in order to arrive at the second operator in (\ref{eq:f7})
we have to replace $U^2_{85} V^1_{57}$ by $U^1_{84} V^2_{47}$ using the $F$-term condition 
for the field $Y_{78}$. Using similar steps it can be shown that the last equation in (\ref{eq:relations3})
indeed holds when we replace the polynomials $f_7$, $b$ and $d$ by the appropriate operators.

Let us represent the corresponding operators for the rest of the polynomials:

\bea  \label{eq:fi}
f_1 \quad &\longleftrightarrow& \quad \textrm{Tr} \, Y_{43} U^1_{35} V^1_{57} U^1_{76} V^1_{68} U^1_{84}, \ldots
\nonumber \\
f_2 \quad &\longleftrightarrow& \quad \textrm{Tr} \, Y_{56} V^2_{63} V^2_{38} U^2_{85}, \ldots
\nonumber \\
f_3 \quad &\longleftrightarrow& \quad \textrm{Tr} \, Y_{43} U^1_{35} Y_{56} V^1_{68} U^1_{84}, \ldots
\nonumber \\
f_4 \quad &\longleftrightarrow& \quad \textrm{Tr} \, Y_{43} Y_{37} U^2_{72} Y_{25} V^2_{54}, \ldots
\nonumber \\
f_5 \quad &\longleftrightarrow& \quad \textrm{Tr} \, U^1_{84} V^2_{47} U^2_{72} Z_{21} U^2_{16} V^2_{63} V^2_{38}, \ldots
\nonumber \\
f_6 \quad &\longleftrightarrow& \quad \textrm{Tr} \, V^2_{38} U^1_{84} U^1_{42} Z_{21} U^1_{13} , \ldots
\eea
Here the dots remind us that in general there are many other operators related to the same polynomial,
which are equivalent by virtue of the $F$-term relations.

Finally, we are in a position to identify two non-holomorphic ``short-cut" operators 
corresponding to the $H=1$ excitation of the string, as we have discussed in the end of the previous section.

Let us focus on the null geodesic that lies at $x=x_2$ and $\theta=\frac{\pi}{2}$. We saw in the previous section 
that for chiral primaries operators the string Hamiltonian takes the following form (see (\ref{eq:HbpsLpqr}) and the
discussion following it):

\be
H_{\textrm{C.P.}} = n_2 + n_4.
\ee
This immediately implies that the polynomial corresponding to the ground state $H=0$ 
is $a$, which in turn is associated with the chiral operator $\mathcal{L}^\uparrow_-$.
Furthermore, the first excited state with $(n_2,n_4)=(0,1)$ is related to the polynomial 
$c$ and the corresponding operator is $\mathcal{L}^\downarrow_-$. 
From (\ref{eq:www}) it follows that for $(n_2,n_4)=(1,0)$
the polynomial is $f_3$ and the relevant operator appears in (\ref{eq:fi}).

Now let us address the construction of the ``short-cut" operators. In the $Y^{p,q}$
example we multiplied the ground state operator by one of the $U(1)$ currents and then removed
a chiral primary operator corresponding to $H=1$. The ``short-cut" operator produced this way is expected to
correspond to $H=1$, since for the current we have $H=2$. 
We will adopt this way of construction also for the case at hand. 
The only novel feature in the $L^{1,7,3}$ case is that we will start from a product of two
operators corresponding to the ground state. This, of course, does not alter the final $H=1$ result 
for the ``short-cut" operator.
For the first operator we have in a schematic way:

\be
U_{35}^1 Y_{56} Y_{64} Y_{43} \, \cdot \, Y_{78} U_{84}^1 Y_{43} Y_{37} \, \cdot \, \bar{U}_{58}^2 U_{85}^2
=
U_{35}^1 \bar{U}_{58}^2 U_{84}^1 Y_{43}  \cdot Y_{56} Y_{64} Y_{43} Y_{37} Y_{78} U^2_{85}.
\ee
Here the first two terms on the left hand side are different representations of the $H=0$ ``long"
operator $\mathcal{L}^\uparrow_-$ and the third term corresponds to the $U(1)$ current.
The last term on the right hand side is the $\mathcal{L}^\downarrow_-$ operator. We therefore conclude
that the ``short-cut" operator we are interested in is:

\be
\mathcal{O}^{(1)} = \textrm{Tr} \, U_{35}^1 \bar{U}_{58}^2 U_{84}^1 Y_{43}.
\ee
Clearly there are many other ``short-cut" operators with exactly the same quantum numbers,
since we could have started with other representations for the operators  $\mathcal{L}^\uparrow_-$
and $\mathcal{L}^\downarrow_-$.

For the second operator we write:
\be
U_{35}^1 Y_{56} Y_{64} Y_{43} \, \cdot \, Y_{37} Y_{78} U_{84}^1 Y_{43}  \, \cdot \, \bar{V}_{86}^1 V_{68}^1
=
\bar{V}_{86}^2 Y_{64} Y_{43} Y_{37} Y_{78} \cdot U^1_{35} Y_{56} V^1_{68} U^1_{84} Y_{43}.
\ee
The first two terms on the left hand side are the ``long" operator $\mathcal{L}^\uparrow_-$
and the third term corresponds to the $U(1)$ current. On the right hand side the second term 
is an operator associated with the polynomial $f_3$ (see (\ref{eq:fi})) and the first term is the
the second ``short-cut" operator:

\be
\mathcal{O}^{(2)} = \textrm{Tr} \, \bar{V}^1_{86} Y_{64} Y_{43} Y_{37} Y_{78}.
\ee
Again, there are many other operators equivalent to these ``short-cut" operators that one can
easily derive starting from different representations for the operators corresponding to the polynomials
$a$ and $f_3$.

%---------------------------------------------------------
\section{Conclusions}
\label{conc}

In this paper we have investigated the Penrose limit 
of the $Y^{p,q}$ and $L^{p,q,r}$ families of Sasaki-Einstein geometries.
The results presented here, therefore, extend the previous studies of 
\cite{Itzhaki:2002kh,Gomis:2002km,PandoZayas:2002rx}. 
In contrast to the Klebanov-Witten model, however, the quiver gauge theories 
dual to the new backgrounds are quite involved, so a straightforward analysis of
the $F$-term relations becomes a formidable task. On the other hand working with
polynomials of the K\"ahler quotient coordinates provides an easy way to identify 
the ground state BMN operator as well as the chiral operators dual to the first excited string state.
We have also given a general idea how to construct non-chiral ``short-cut" operators of \cite{Benvenuti:2005cz} 
in the $L^{p,q,r}$ models and provided an explicit solution in the $L^{1,7,3}$ case.

Unfortunately, we were not able to perform the supergravity spectrum analysis in these backgrounds,
in order to verify that the ``short-cut" operators have proper dimensions matching the 
first excited string state.
The first step towards this direction was done \cite{Kihara:2005nt,Oota:2005mr}
and it will be very interesting to pursue this direction in the future.

%===============
\acknowledgments
%===============

We would like to thank Sergio Benvenuti, Leo Pando Zayas, Ami Hanany, Yaron Oz, Ricardo Argurio, Daniel Persson  
for fruitful discussions. The work of O. Mintkevich and J. Sonnenschein 
was partially supported by the Israel Science Fundation and by a grant of 
German-Israeli Project Cooperation - DIP Program (DIP H.52).

\bibliography{ppYpq}

\end{document}